\documentclass[twocolumn,showpacs,floatfix,prb]{revtex4-1}
\usepackage{graphicx}
\usepackage{float}
\usepackage{dcolumn}
\usepackage{bm}
\usepackage{amssymb}
\usepackage{amsmath}
\usepackage{hyperref} 
\hypersetup{bookmarks=true,unicode=true,colorlinks=true, urlcolor=blue,linkcolor=blue,citecolor=blue,filecolor=blue}

\usepackage{physics}

\begin{document}

\title{Evolution of topological superconductivity by orbital selective \\ 
confinement in oxide nanowires}

\author{C.A. Perroni, V. Cataudella} 
\address{CNR-SPIN and Physics Department "Ettore Pancini", 
Universita' degli Studi di Napoli Federico II,
\\ Complesso Universitario Monte S. Angelo, Via Cintia, I-80126 Napoli, Italy}
\author{M. Salluzzo} 
\address{CNR-SPIN,
 Complesso Universitario Monte S. Angelo, Via Cintia, I-80126 Napoli, Italy}
\author{M. Cuoco, R. Citro}
\address{CNR-SPIN and Physics Department "E.R. Caianiello", 
Universita' degli Studi di Salerno,\\
Via Giovanni Paolo II, 132, I-84084 Fisciano (Sa), Italy }

\begin{abstract}
We determine the optimal conditions to achieve topological superconducting phases having spin-singlet pairing for a planar nanowire with finite lateral width in the presence of an in-plane external magnetic field. We employ a microscopic description that is based on a three-band electronic model including both the atomic spin-orbit coupling and the inversion asymmetric potential at the interface between oxide band-gap insulators. We consider amplitudes of the pairing gap, spin-orbit interactions and electronic parameters that are directly applicable to nanowires of LaAlO$_3$-SrTiO$_3$. 
The lateral confinement introduces a splitting of the $d$-orbitals that alters the orbital energy hierarchy and significantly affects the electron filling dependence of the topological phase diagram. Due to the orbital directionality of the $t_{2g}$-states, we find that in the regime of strong confinement the onset of topological phases is pinned at electron filling where the quasi flat heavy bands start to get populated. The increase of the nanowire thickness leads to a changeover from sparse-to-dense distribution of topologically non-trivial domains which occurs at the cross-over associated to the orbital population inversion.
These findings are corroborated by a detailed analysis of the most favorable topological superconducting phases in the electron doping-magnetic field plane highlighting the role of orbital selective confinement.
\end{abstract}

\maketitle

\section{Introduction}
In the last years there has been a growing interest in topological superconducting phases both for fundamental perspectives in theoretical physics and the potential impact for achieving novel devices towards emergent technologies \cite{Book1,Book2,RevZhang,RevSato1}. Topological superconductors host new states of matter exhibiting topological order and topologically nontrivial structures of electron pairing. Remarkably, boundaries or defects of topological superconductors can lead to zero energy Majorana modes, whose formation and stability is ensured by the topological non-trivial character of the ground-state in the bulk \cite{RevSato2}. A key model for capturing the fundamental features of topological superconductors is represented by the spinless p-wave Kitaev model \cite{RevSato1,RevSato2,Kitaev}.
Apart from the non-standard physical phenomena arising in materials platforms that host Majorana zero modes, due to their non-Abelian character these materials have been indicated as fundamental building blocks for designing systems in the field of topological quantum computation \cite{Kitaev,QuantumSarma}.

Seminal theoretical proposals mostly focused on hybrid supercondutor-semiconductor nanowire junctions as promising platforms to achieve topological superconductors and design Majorana bound states \cite{Sau1,Lutchyn1,Oreg1,Aguado1}. Signatures of topological superconducting phases have been then experimentally investigated in these heterostructures, using InSb and InAs semiconductors \cite{Mourik,Albrecht,Deng,Zhang,Lutchyn2}.
Concerning the design of Majorana states there are three key elements which are recognized to be fundamentally relevant. A first crucial ingredient to generate topological superconductivity is the use of non-centrosymmetric materials with large Rashba spin-orbit coupling, which results into the removal of the spin degeneracy of the electronic bands in momentum space. The second ingredient is the presence of time-reversal symmetry breaking magnetic field, which allows to describe the system by effective spinless degrees of freedom. The third one is superconducting pairing acting on those electronic states. For instance, superconducting phases can be obtained by the 
proximity with a standard spin-singlet superconductor.  The pairing interaction binds electrons which acquire a gap at all the momenta in the reciprocal space. The transition from trivial to topological superconducting state then typically occurs at finite amplitude of the applied magnetic field.

Majorana edge modes have been predicted for nanowires proximity-coupled with an s-wave superconductor not only in the case of a single electronic band, but also for multiple sub-bands due to the lateral confinement of the nanowire \cite{Wimmer,Potter,Lutchyn3,Lim,Stanescu} and networks \cite{Alicea}. Indeed, in these systems, a nontrivial topological state can be realized when an odd number of sub-bands is occupied (corresponding to a odd number of Fermi point pairs). Moreover, the occupation of multiple bands  may enhance the stability of superconducting phases against disorder by increasing the carrier density.  However, the inter-band mixing due to the Rashba spin-orbit term hybridizes Majorana pairs originating from different transverse modes \cite{Bena}, which is against the stability of the topological phase.

Additional materials and platforms have been further considered with the aim to search for topological superconductivity (TSC).  One strategy has been to introduce dopant into a topological insulator \cite{Wray}, or to exploit the surface states of iron-based superconductors\cite{Zhang2}.  Recently, the quasi-2D electron gases (2DEGs) formed at the interface between $LaAlO_3$ and $SrTiO_3$ (LAO/STO) \cite{Ohtomo} have been theoretically proposed as possible candidates for the realization of topological superconducting phases in two-dimensional \cite{Scheurer,Mohanta,Loder,Fukaya} and effective quasi one-dimensional  models \cite{Fidkowski1,Fidkowski2,Mazziotti}.
Oxide q2DEGs, indeed, are characterized  by the simultaneous presence of strong spin-orbit coupling \cite{Caviglia} and  superconductivity \cite{Reyren}, both widely tunable by electric field effect \cite{Caviglia,Caviglia2}, while 2D magnetism, coexisting with superconductivity, \cite{Hwang,Pai} can be induced by opportune atomic engineering of the heterostructures \cite{Stornaiuolo}.  

In this paper, we aim to investigate the emergence of a topological superconducting phase in oxide nanowires with variable lateral width. We employ a microscopic description that is well suited for the low-energy electronic states of the Ti $t_{2g}$ orbitals close to the Fermi level and includes the atomic spin-orbit and inversion asymmetric potential associated with an orbital Rashba interaction. In addition, we consider a magnetic field lying into the plane of the LAO/STO interface as a source of time reversal symmetry breaking and depairing, assuming a conventional intra-orbital spin-singlet s-wave superconductivity.  Due to the presence of multi-bands and their reciprocal coupling, the electronic spectrum is more intricate than that obtained in the presence of spin-Rashba coupling for quasi one-dimensional nanowires \cite{Bena,Perroni1}.
Indeed, we focus on the role played by the lateral confinement emphasizing the distinctive marks of the orbital degrees of freedom associated with the Ti $t_{2g}$ orbitals. The finite lateral size of the nanowire introduces a splitting of the $d$-orbitals that alters the orbital energy hierarchy and significantly affects the electron filling dependence of the topological phase diagram. Due to the orbital directionality of the $t_{2g}$-states, we find that in the regime of strong confinement the onset of topological phases is pinned close to electron filling where the quasi flat ($yz$) heavy bands start to get populated. The increase of the nanowire thickness, then, is able to drive a changeover from sparse-to-dense distribution of topologically non-trivial domains which occurs at the cross-over associated to the orbital population inversion. 
These findings are achieved by a detailed analysis of the most favorable topological superconducting phases based on self-consistent computation of the order parameter within the nanowire in the parameters space set by the electron filling and the amplitude of the magnetic field. The investigation highlights the emergent role of orbital selective confinement in a microscopic regime with amplitudes of the pairing gap, spin-orbit interactions and electronic parameters that are directly applicable to nanowires of LAO/STO.

The paper is divided as follows. In Section 2, the model hamiltonian for nanowires at the LAO/STO interface is introduced and the methodology for the determination of the topologically stable superconducting phase is presented. 
In Section 3, we provide a comparative study of the electronic structure of the nanowire for different lateral confinements and we present the most relevant aspects of the topological phase diagram. Section 4 is devoted to the discussion and the concluding remarks  by also dealing with the comparison between the properties of LAO/STO nanowires and hybrid supercondutor-semiconductor heterostructures employing semiconducting nanowires like InSb and InAs. Additional results of the topological phase diagram for the single chain nanowire are reported in Appendix \ref{AppA}, while details about the calculation of the topological invariant are in Appendix \ref{AppB}.

\section{Model and methodology} 

To introduce the model of the examined nanowires, we start by considering a confined two-dimensional (2D) electronic system  with broken out-of-plane inversion symmetry and having only the $t_{2g}$-orbitals close to the Fermi level. Due to the weak octahedral distortions, the transition metal (TM)-oxygen(O) bond angle is almost ideal and thus the three $t_{2g}$-bands are mainly directional and basically decoupled, e.g., an electron in the $d_{xy}$-orbital can predominantly hop along the $y$ or $x$ direction through the intermediate $p_x$ or $p_y$-orbitals. Similarly, the $d_{yz}$ and $d_{zx}$-bands are quasi one-dimensional when considering a 2D TM-O bonding network. Such directional character of the $t_{2g}$ is very relevant when one is also including the effects of a lateral in-plane confinement as the confining potential acts in a different way on the three $t_{2g}$ orbitals. 

Concerning the inversion asymmetry, we consider microscopic couplings that arise from the out-of-plane oxygen displacements around the TM. 
Indeed, by breaking the reflection symmetry with respect to the plane placed in between the TM-O bond, a mixing of orbitals that are even and odd under such a transformation is generated. 
Such crystal distortions are very important in 2D electron gas forming at the interface of insulating polar and non-polar oxide materials or on their surface and they result in the activation of an effective hybridization, which is odd in space, of $d_{xy}$ and $d_{yz}$ or $d_{zx}$-orbitals along the $y$ or $x$ directions, respectively. Such type of interaction is commonly dubbed orbital Rashba coupling. Indeed, the inversion symmetry breaking is primarily affecting the orbital degrees of freedom and then it is the atomic spin-orbit that transfers the inversion asymmetric potential into the spin-sector.
The atomic spin-orbit interaction (SO) is then a crucial term to be included into the electronic description and mixes the spin-orbital degrees of the $t_{2g}$-states thus competing with the quenching of the orbital angular momentum due to the crystal potential.
Since we are interested in the consequences of time reversal symmetry breaking due to an external magnetic field and in the conditions to achieve a topological non-trivial superconducting phase, the model Hamiltonian is also including a Zeeman coupling between the magnetic field and the spin of the carriers.  

Thus, the model Hamiltonian, including the $t_{2g}$ hopping connectivity, the atomic spin-orbit coupling, the inversion symmetry breaking term, and the external magnetic field can be expressed as follows \cite{Zhong,Khalsa,Vivek}
\begin{align}
\mathcal{H}&=\sum_{\bm{k}}\Hat{D}(\bm{k})^{\dagger}H(\bm{k})\Hat{D}(\bm{k}), \\
H(\bm{k})&=H^0+H^\mathrm{SO}+H^{Z}+H^{M},
\end{align}%
where $\Hat{D}^{\dagger}(\bm{k})=\left[ c^{\dagger}_{yz\uparrow \bm{k}}, c^{\dagger}_{zx\uparrow \bm{k}}, c^{\dagger}_{xy\uparrow \bm{k}}, c^{\dagger}_{yz\downarrow \bm{k}}, c^{\dagger}_{zx\downarrow \bm{k}}, c^{\dagger}_{xy\downarrow \bm{k}} \right]$ is a vector whose components are associated with the electron creation operators for a given spin $\sigma$ ($\sigma=[\uparrow,\downarrow]$), orbital $\alpha$ ($\alpha=[xy,yz,zx]$), and momentum $\bm{k}$ in the Brillouin zone.

In order to write down the various terms of the Hamiltonian, it is convenient to introduce the matrices $\Hat{l}_{x}$, $\Hat{l}_{y}$ and $\Hat{l}_{z}$, which are the projections of the $L=2$ angular momentum operator onto the $t_{2g}$ subspace, i.e.,
\begin{align} 
\Hat{l}_{x}&=
\begin{pmatrix}
0 & 0 & 0 \\
0 & 0 & i \\
0 & -i & 0
\end{pmatrix}, \\
\Hat{l}_{y}&=
\begin{pmatrix}
0 & 0 & -i \\
0 & 0 & 0 \\
i & 0 & 0
\end{pmatrix}, \\
\Hat{l}_{z}&=
\begin{pmatrix}
0 & i & 0 \\
-i & 0 & 0 \\
0 & 0 & 0
\end{pmatrix},
\end{align}%
assuming $\{d_{yz}, d_{zx}, d_{xy}\}$ as orbital basis.

Then, including also the spin degrees of freedom, in the spin-orbital basis, $H_0(\bm{k})$ is given by
\begin{align}
&H^0=\Hat{\varepsilon}_{\bm{k}} \otimes \Hat{\sigma}_{0}, \\
&\Hat{\varepsilon}_{\bm{k}}=
\begin{pmatrix}
\varepsilon_{yz} &0 &0 \\
0 & \varepsilon_{zx} &0 \\
0 &0& \varepsilon_{xy}
\end{pmatrix}, \notag \\
&\varepsilon_{yz}=2t_{1y}\left( 1-\cos{k_y}\right)+2t_{2x}\left(1-\cos{k_x}\right), \notag \\
&\varepsilon_{zx}=2t_{1x}\left(1-\cos{k_x}\right)+2t_{2y}\left(1-\cos{k_y}\right), \notag \\
&\varepsilon_{xy}=4t_{1}-2t_{1x}\cos{k_x}-2t_{1y}\cos{k_y}+\Delta_{t}, \notag
\end{align}%
where $\Hat{\sigma}_{0}$ is the unit matrix in spin space. 
Here, we assume that $t_{1x}=t_{1y}=t_1$ and $t_{2x}=t_{2y}=t_2$ are the orbital dependent hopping amplitudes. 
$\Delta_{t}$ denotes the crystal field potential as due to the symmetry lowering from cubic to tetragonal also related to inequivalent in-plane and out-of-plane transition metal-oxygen bond lengths.
The symmetry reduction yields a level splitting between $d_{xy}$-orbital and $d_{yz}/d_{zx}$-orbitals.
$H^\mathrm{SO}$ denotes the atomic $\bm{L} \cdot \bm{S}$ spin-orbit coupling,
\begin{align}
H^\mathrm{SO}
=\Delta_{\mathrm{SO}}\left[ \Hat{l}_x \otimes \Hat{\sigma}_x+\Hat{l}_y \otimes \Hat{\sigma}_y+\Hat{l}_z \otimes \Hat{\sigma}_z \right], 
\end{align}%
with $\Hat{\sigma}_{i}(i=x,y,z)$ being the Pauli matrix in spin space.

As mentioned above, the breaking of the mirror plane in between the TM-O bond, due to the out-of-plane oxygen displacements,
yields an inversion symmetry breaking term $H^{Z}(\bm{k})$ of the type
\begin{align}
H^{Z}
=\gamma \left[ \Hat{l}_y \otimes \Hat{\sigma}_0 \sin{k_x}-\Hat{l}_x \otimes \Hat{\sigma}_{0} \sin{k_y} \right].
\end{align}%
This contribution gives an inter-orbital process, due to the broken inversion symmetry, that mixes $d_{xy}$ and $d_{yz}$ or $d_{zx}$ along $x$ or $y$ spatial directions.

Finally, we consider the effects of a magnetic field lying into the plane of the LAO/STO interface. The resulting Zeeman coupling is described by the term $H^{M}$, which indicates the interaction of an external homogeneous magnetic field to the electron spins for each orbital flavor

\begin{align}
H^\mathrm{M}
=\left[ M_x \Hat{l}_0 \otimes \Hat{\sigma}_x+M_y \Hat{l}_0 \otimes \Hat{\sigma}_y \right], 
\end{align}%

\noindent with $\Hat{l}_0$ being the unit matrix in the orbital space. Therefore, the magnetization examined in this work is non local and  induced only by the external magnetic field.
 We remark that we are analyzing the clean limit of the system, hence we are not considering the coupling of itinerant electrons  with magnetic impurities \cite{Fidkowski1}.   Moreover, due to the spin-orbit coupling, the presence of a magnetization in the spin channel sets also an orbital polarization. Since the spin-orbit coupling is typically  larger than the strength of the applied magnetic field, the inclusion of the orbital coupling to the field will be a correction \cite{Ilaninew}.

\begin{figure}[t]
\centering
\includegraphics[width=8.5cm,height=7.0cm]{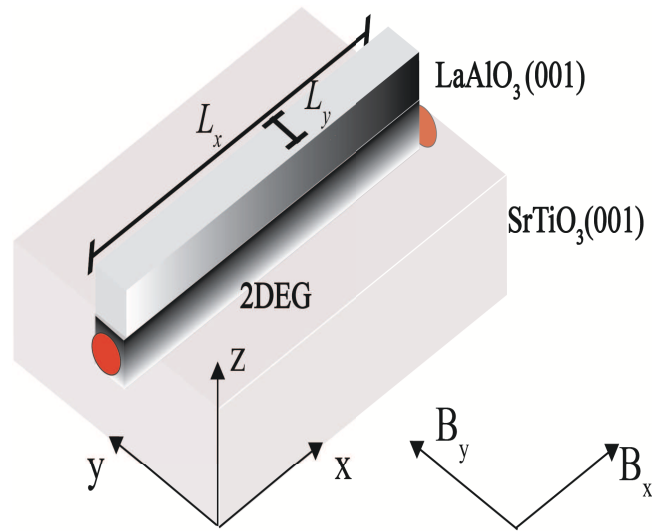}
\caption{Sketch of the nanowire at the LAO/STO interface. The nanowire length is indicated by $L_x$, its width by $L_y$. The quantities $B_x$ and $B_y$ indicate the cartesian components of the in-plane magnetic field. The red points highlight the Majorana edge modes characteristic of the topological superconducting phases.}
\label{FigDev}
\end{figure}

Concerning the superconducting pairing, we assume that the interaction is local, with spin-singlet symmetry and active only for electrons sharing the same orbital symmetry.
Hence, the superconducting term $H^P$ can be expressed as
\begin{equation}
H^P = -U \sum_{{\bf{i}},\alpha}   n_{{\bf{i}}\alpha,\uparrow} n_{{\bf{i}}\alpha,\downarrow},
\label{hamilpair}
\end{equation}
where $U$ is the pairing energy, and $n_{{\bf{i}}\alpha,\sigma} =c_{{\bf{i}},\alpha,\sigma}^{\dagger} c_{{\bf{i}}\alpha,\sigma}$ is the local spin-density operator for the $\sigma$ polarization and the $\alpha$ orbital, at a given position ${\bf{i}}=(i_x,i_y)$ in the square lattice (with parameter $a=3.9 \AA$), with $i_x$ and $i_y$ the corresponding coordinates along the $x$ and $y$ symmetry axes.
Since we are interested in the spatial profile of the superconducting order parameter for the case of a nanostrip with a finite lateral thickness in the $xy$ plane, it is useful to introduce the superconducting order parameter in the real space. 
In order to get the Bogoliubov-de Gennes equations, we then employ the usual decoupling scheme for the pairing term using a mean-field approach for the spatial and orbital degrees of freedom:
\begin{eqnarray}
H^P & \approx&  - \sum_{{\bf{i}},\alpha} \Delta_{{\bf{i}},\alpha} \left[ c_{{\bf{i}},\alpha,\uparrow}^{\dagger} c_{{\bf{i}},\alpha,\downarrow}^{\dagger} + c_{{\bf{i}},\alpha,\downarrow}  c_{{\bf{i}},\alpha,\uparrow}  \right]  \nonumber \\
&& + U \sum_{{\bf{i}},\alpha} D^2_{{\bf{i}},\alpha},
\label{hamilbogo}
\end{eqnarray}
with the pairing amplitude $D_{{\bf{i}},\alpha}=\langle c_{{\bf{i}},\alpha,\downarrow}  c_{{\bf{i}},\alpha,\uparrow}\rangle$ and the order parameter $ \Delta_{{\bf{i}},\alpha}=U D_{{\bf{i}},\alpha}$ are taken in a gauge such as to have a real amplitude. Here, $\langle A \rangle$ stands for the ground state average of any given operator $A$. The solutions due to this decoupling corresponds to a local s-wave pairing which is considered to be one of the most favored superconducting instability in the two-dimensional bulk. \cite{Michaeli,Nakamura,Loder1,Mohanta1} To assess the stability of the superconducting states in the presence of an applied magnetic field, we solve the Bogoliubov-de Gennes equations in a self-consistent way by an iterative scheme of computation until the desired accuracy is achieved. We point out that, apart from indications arising through the determination of the topological invariant, the quantum transition from trivial to topological superconducting states is also signaled by a rapid decrease of the order parameter as a function of the magnetic energy ($M_x$ or $M_y$).
Considering that in the two-dimensional bulk the order parameter $\Delta$ is of the order of $0.1$ meV, one can derive a lower bound for the superconducting coherence length  $\xi$:
 \begin{equation}
\xi \simeq \frac{a t_2}{\Delta} \simeq 200 a \simeq 80 nm,
\label{coherence}
\end{equation}
which is in good agreement with experimental estimates \cite{Pai}.
Assuming the structure of the model Hamiltonian for the uniform two dimensional case, it is straightforward to obtain the description for a nanowire with finite thickness along one of the crystal symmetry directions (see Fig. (\ref{FigDev}) for a sketch of the nanowire at the LAO/STO interface).

Before starting with the discussion of the results, it is useful to set the energy scales for the various terms of the Hamiltonian taking into account the targeted materials.
Concerning the hopping amplitudes, we assume that $t_1=300$ meV and $t_2=20$ meV \cite{Pai,Zhong,Joshua}. These values in the limit of small momentum would lead to a light mass $m_1$, associated with the hopping $t_1$, that turns out to be smaller than the free electron mass $m_e$ 
\begin{equation}
m_1=\frac{\hbar^2}{2t_1 a^2} \simeq 0.7 m_e
\label{massa1}
\end{equation}
and a heavy mass $m_2$ corresponding to the hopping amplitude $t_2$
\begin{equation}
m_2=\frac{\hbar^2}{2t_2 a^2} \simeq 10 m_e.
\label{massa2}
\end{equation}
\noindent with $\hbar$ being the Planck constant.

The energies of the various orbitals in the $t_{2g}$ sector at the $\Gamma$ point are assumed to be split by the crystal field potential $\Delta_t$ that takes into account the lowering of the symmetry from cubic to tetragonal. According to the ab-initio estimates and on the basis of spectroscopic studies \cite{Pai,Zhong,Joshua}, the crystal field potential is larger than the spin-orbit coupling and the inversion asymmetric potential. Here, we assume that the bare $\Delta_t=-50$ meV. Slight variations of the electronic parameters do not alter the qualitative outcomes of the achieved results.  
Concerning the interaction due to the breaking of the inversion symmetry, the strength of $\gamma$ is assumed to be 20 meV \cite{Zhong}. The atomic spin-orbit coupling $ \Delta_{SO}$ is taken to be 10 meV \cite{Zhong} according to the typical estimates employed for the Ti element. 

Concerning the value of the orbital Rashba coupling one can observe that for electron filling corresponding to the unique occupation of the $xy$ band, due to the crystal field splitting, one can derive an effective spin Rashba interaction $\alpha_R$ by performing a perturbation theory to the second order in the spin-orbital interactions \cite{Zhong}. Hence, using the parameter values fixed in the previous paragraph, one gets that a rough estimate of $\alpha_R$ is given by 
\begin{equation}
\alpha_R \simeq  \frac {a \gamma \Delta_{SO}} {|\Delta_t|}  \simeq 1.6 \  meV \cdot  nm,
\label{alfar}
\end{equation}
a value compatible with experimental measurements \cite{Caviglia,Pai} in the limit of low values of particle density. 

Finally, in order to assess the topological character of the superconducting phase for the nanowire we observe that, due to the applied magnetic field, the time reversal symmetry is broken in the examined nanowires. Thus, according to the Altland-Zirnbauer classification \cite{Book1}, the Hamiltonian is in the $D$ symmetry class, and, since the local spin-singlet pairing makes the spectrum fully gapped, one can introduce the following $Z_2$ topological invariant, whose expression for translational invariance along the $x$ direction is given by
\begin{equation}
Q =sgn \left[ P(k_x=0) P\left( k_x=\frac{\pi}{a} \right) \right],
\label{invariant}
\end{equation}
where $sgn$ indicates the signum function and $P(k_x)$ is the Pfaffian of the skew-symmetric matrix derived from the Bogoliubov-de Gennes Hamiltonian upon a transformation in the Majorana basis evaluated in the particle-hole symmetric points, i.e. at $k_x=0$ and $k_x=\pi/a$ \cite{Kitaev}. The value $Q=1$ marks a trivial superconducting state, while $Q=-1$ a topological state.  Details about the calculation of the topological invariant are provided in Appendix \ref{AppB}.

\section{Electronic structure and stability of topological superconductivity: role of in-plane confinement}


The key aim of the present analysis is to determine the optimal conditions for setting a topological  superconducting phase upon the application of an applied magnetic field assuming that the electrons are confined in a planar oxide nanowire with a finite width along one of the crystal symmetry axis.  
The geometry of the nanostrip exhibits a confinement due to the finite tickness in the $y$ direction corresponding with a length $L_y=N_y a$, where $N_y$ denotes the number of chains. For mimicking the inhomogeneous effects of the confining potential, we use open boundary conditions (OBC) along the $y$ axis with hard wall confinement at $y=0$ and $y=L_y$. On the other hand, along the longitudinal $x$ direction of the nanowire, translational invariance and periodic boundary conditions (PBC) are assumed when dealing with the determination of the topological phase diagram. 

\begin{figure}[t]
\centering
\includegraphics[width=10.0cm,height=9.0cm,angle=-90]{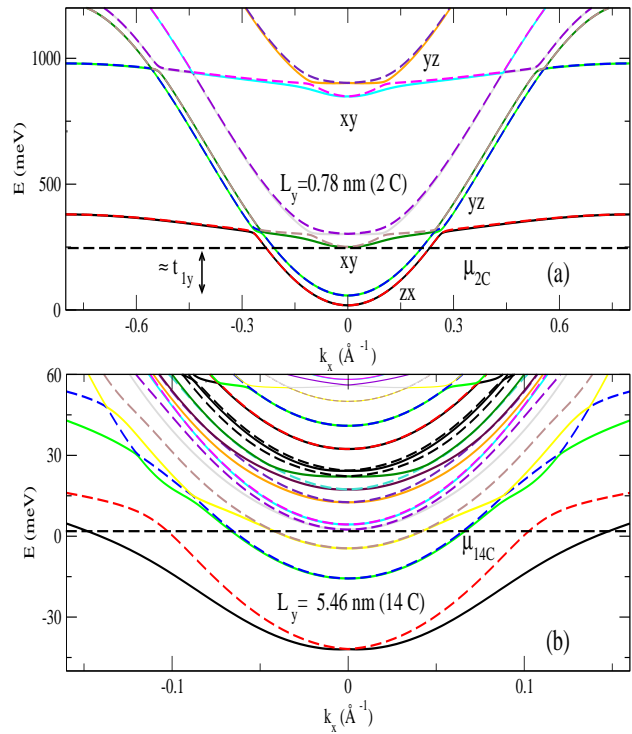}
\caption{Electronic structure of the oxide nanowire for the case of (a) 2-chains (2C), and (b) 14-chains (14C) thickness. For the 2C nanostrip, the effect of the confinement is to push the $xy$ and $yz$ bands above in energy with respect to the $zx$ state.}
\label{Fig1}
\end{figure}

We start by considering how the electronic structure in the normal state is modified by changing the number of chains of the nanowire and thus, effectively, the strength of the confining potential along the transverse direction. 
In Fig. \ref{Fig1}, we report the zero-field electronic states for two different configurations that are marked by an inequivalent confining potential. In the case of only one chain, discussed in Appendix \ref{AppA}, and in the case of two-dimensional bulk \cite{Pai}, the $xy$ band is the lowest in energy and it exhibits a splitting of the order of $\Delta_t$ with the $xz,yz$ states, that are in turn separated in energy as a consequence of the atomic spin-orbit coupling $\Delta_{SO}$.


\begin{figure}[t]
\centering
\includegraphics[width=9.5cm,height=9.0cm,angle=-90]{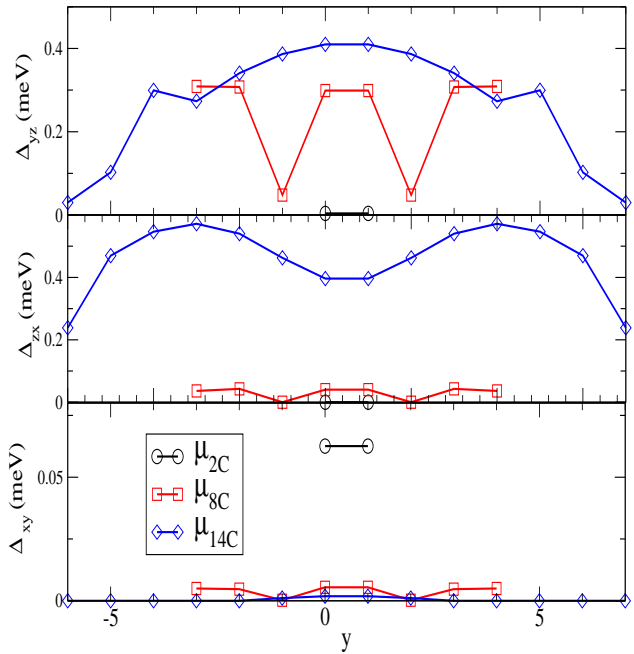}
\caption{(a) Superconducting order parameters for $yz$-like bands (upper panel), $zx$-like  bands (middle panel), and $xy$-like bands (lower panel) as a function of the position y of the lateral chains for different wire widths: black circles for $N_y=2$, red squares for $N_y=8$, blue diamonds for $N_y=14$. The chemical potentials $\mu_{2C}$ for $N_y=2$ and $\mu_{14C}$ for $N_y=14$ are indicated in Fig.(\ref{Fig1}). } 
\label{Fig2}
\end{figure}

Considering the nanostrip with $N_y=2$ (indicated as 2C) leads to a remarkable rearrangement of the energy spectrum (see Fig. \ref{Fig1}(a)). 
The zero-field spectrum for the $N_y=2$ presents six doublets which are quite close in energy (around 1 eV). The lowest two doublets do not show any sizable effective Rashba splitting for small values of momentum $k_x \sim 0$ because they are coupled by second order terms whose amplitude is a fraction of meV as due to processes that involve the inter-chain splitting, i.e. $\sim \frac{\Delta_{SO} \gamma}{t_1}$. The highest energy configurations present a more pronounced double minimum structure that is typical of an effective spin Rashba coupling.  In particular, since the $xz$ bands are not so sensitive to the confinement along the $y$ direction, they correspond to the low energy doublet (Fig. \ref{Fig1} (a)).  On the other hand,  the $xy$ and $yz$ bands (third/fifth and fourth/sixth, respectively) are strongly influenced by the confinement along $y$-direction which pushes them at higher energies. Specifically, the $zx$ bands are lowered with respect to the $xy$ band by an energy separation of the order of $t_1$ (i.e. $\sim 300$ meV).

As discussed above, one can clearly visualize the effect of the confinement on the bands $xy$ and $yz$ which have a large hopping amplitude along the transverse direction. Such orbital inversion with a lowering of the ${zx}$ band holds for thickness amplitudes that are below $N_y=10$. Above this threshold the dispersion of the $d_{xy}$ bands along the $y$ direction allows again to make them the lowest occupied in energy and the corresponding sub-bands tend to reconstruct the electronic spectrum of the two-dimensional bulk limit. 

To directly visualize such changeover of the sub-bands, in Fig. \ref{Fig1}(b) we report the zero field band-structure for a representative length of intermediate confinement, i.e. $N_y=14$ (indicated as 14C). At low energy, the first couples of bands have $d_{xy}$ origin and show a conventional spin-Rashba effective coupling. However, with increasing energy, it is not immediate to distinguish the character of the electronic states, since higher sub-bands with 
$xy$ character become closer in energy with the lower sub-bands having $yz$ and $zx$ orbital content and get mixed by the orbital Rashba and the spin-orbit coupling. Indeed, in this regime, all the bands become strongly interrelated with increasing energy and thus through the electron doping too. 



\begin{figure}[t]
\centering
\includegraphics[width=10.0cm,height=9.2cm,angle=-90]{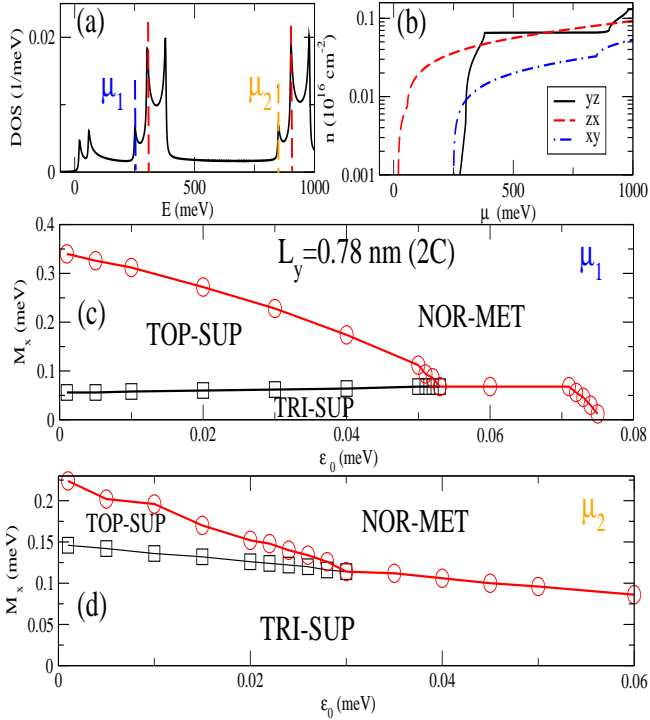}
\caption{(a) Normal state density of states for the $N_y=2$ nanowire. Dotted lines schematically indicate the filling regimes for which a topological superconducting phase can be stabilized. (b) Evolution of the orbital dependent electron density as a function of the energy. The lowest bands have a $xz$ character upon reaching a value of the chemical potential of the order of 250 meV whereas the $xy$ and $yz$ electronic states start to get populated.  (c) Phase diagram at a given value of the band minimum chemical potential ($\mu_1=\mu_0$ defined in Eq. (\ref{mupotent})) in terms of the electron filling offset $\epsilon_0$ and the applied magnetic field $M_x$ along the $x$ direction of the nanowire. TOP-SUP, TRI-SUP and NOR-MET stand for topological superconducting, trivial superconducting, and normal metallic phase, respectively. (d) as in (c) but for a value of the band minimum chemical potential which is given by $\mu_2=\mu_0$ defined in Eq. (\ref{mupotent}) as indicated in the panel (a).} 
\label{Fig3}
\end{figure}

At this point, it is useful and instructive to address the stability of the superconducting phases with increasing the number $N_y$ of lateral chains in the absence of an applied magnetic field.  In Fig. (\ref{Fig2}), we report the
superconducting order parameters for $yz$-like, $zx$-like, and $xy$-like bands as a function of the position y of the lateral chains. In the case $N_y=2$, we consider the superconducting phases  corresponding to the chemical potential  $\mu_{2C}$ shown in Fig. (\ref{Fig1}) (in Fig. (\ref{Fig3}) it will be coincident with $\mu_1$). As discussed above, the less occupied band is mainly $xy$-like, therefore, one expects that the superconducting instabilities mostly involve this kind of orbitals (at $\mu_1$ there is a peak of the density of states).  Moreover, as reported in Appendix A, the order parameters necessary to stabilize the superconducting phase are of the same order of magnitude of those of the single-chain wire ($U$ of the order of $60$ meV is used). 

\begin{figure}[t]
\centering
\includegraphics[width=10.0cm,height=8.8cm,angle=-90]{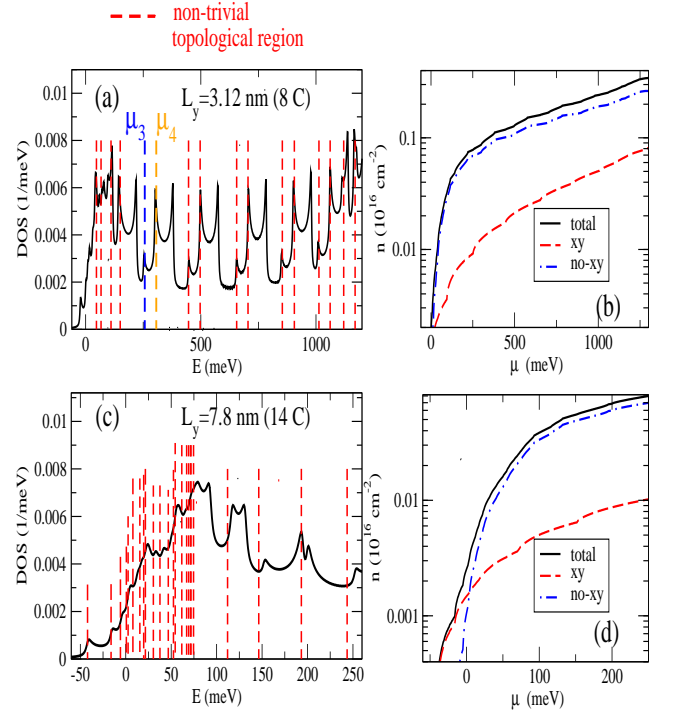}
\caption{(a) Normal state density of states for a nanowire with $N_y=8$. Dotted lines (red) schematically indicate the electron doping regimes for which a topological superconducting phase can be stabilized. (b) Evolution of the orbital dependent and total electron density as a function of the energy. The lowest bands have a dominant $xz$ while the occupation of the $xy$ and $yz$ electronic states have a different compressibility as measured by the derivative of the electron density with respect to the chemical potential.  (c) as in the panel (a) but for  a nanowire with $N_y=14$. (d) electron density vs energy. The low energy electron density is dominated by $xy$ bands. An orbital inversion is obtained at a value of the chemical potential that is $\sim 0$ meV.}
\label{Fig4}
\end{figure}

\begin{figure}[t]
\centering
\includegraphics[width=10.0cm,height=8.9cm,angle=-90]{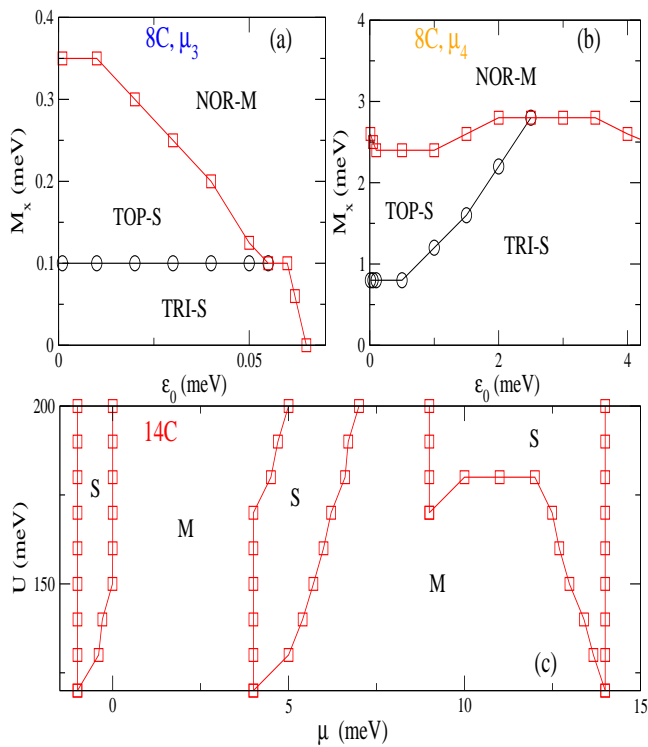}
\caption{ (a) Phase diagram at a given value of the band minimum chemical potential ($\mu_3=\mu_0$ defined in Eq. (\ref{mupotent})) in terms of the electron filling offset $\epsilon_0$ and the applied magnetic field $M_x$ along the $x$ direction of the nanowire for $N_y=8$. TOP-S, TRI-S and NOR-M stand for topological superconducting, trivial superconducting, and normal metallic phase, respectively. (b) as in (a) but for a value of the band minimum chemical potential which is given by $\mu_4=\mu_0$ defined in Eq. (\ref{mupotent}) as indicated in the panel (a).
(c) Phase diagram in terms of the pairing energy $U$ and the chemical potential $\mu$ for $N_y=14$. S stands for superconductor, M for metal, respectively.}
\label{Fig5}
\end{figure}

Next, we analyze the case $N_y=8$ for a chemical potential corresponding to the occupation of the first $yz$-like band ($\mu_{8C}$ will correspond to  the chemical potential $\mu_4$ shown in Fig. (\ref{Fig4}) and (\ref{Fig5})). As reported in middle panel of Fig. (\ref{Fig2}), the largest order parameters are only related  to $yz$ character ($U$ of the order of $100$ meV is used).  We also notice that the spatial behavior of the order parameters is nearly oscillating, therefore it is quite dependent on the lateral boundary conditions. Finally, we study  the case $N_y=14$ determining the spatial profile of  the order parameters  for the chemical potential  $\mu_{14C}$ shown in Fig. (\ref{Fig1}) (U larger than $120$ meV is adopted). For the chemical potential $\mu_{14C}$ close to $0$ meV, the highest subbands shown in the lower panel of Fig. (\ref{Fig2}) present a mixed $yz$/$zx$ character, therefore the order parameters are expected to be magnified for these orbitals. Only in the center of the wire ($y$ close to zero), the order parameters $\Delta_{yz}$ and $\Delta_{zx}$ have similar magnitude. Indeed,  these two order parameters show a different behavior at the boundaries.  Ultimately, the orbital character of the bands is able to affect the behavior of the superconducting order parameters, in particular their spatial profile.

We are ready  to analyze the parameters regime for which a topological superconductivity is achieved in the presence of in-plane magnetic field. We start form the case $N_y=2$. From the analysis of the normal state spectrum and on the basis of the correspondence between the Kitaev model and that one described by a Rashba interaction in the presence of an applied magnetic field, one expects to find a topological superconducting state for values of the electron filling that correspond to a chemical potential located  nearby the minimum of the bands belonging to the third up to the sixth doublet. Here, due to the multi-orbital electronic spectrum, it is convenient to decompose the chemical potential $\mu$ as 
\begin{equation}
\mu =\mu_0 + \epsilon_0,
\label{mupotent}
\end{equation} 
where $\mu_0$ typically indicates the energy minimum of a given band, while $\epsilon_0$ provides the effective energy offset that tunes the filling of the corresponding state. 
Apart from the analysis of the topological invariant introduced in Sect. II, the TSC phase diagram can be also traced out by looking at a sharp variation of the order parameter when a magnetic field is switched on along the longitudinal direction ($M_x \ne 0$) as well as by direct inspection of the band gap closing at $k_x=0$. In Fig. \ref{Fig2} we report the topological phase diagram for $N_y=2$ obtained by varying the filling energy $\epsilon_0$ for two different values of the chemical potential ($\mu_0=\mu_1=250.356$ meV and $\mu_0=\mu_2=848.490$ meV) corresponding to the bottom of the third and fifth doublet of the bands discussed above in Fig.\ref{Fig1}. We notice that to get a topological superconductivity in the upper band of $xy$ character (see Fig. \ref{Fig2}(d)) higher magnetic fields are required. For the third pair of doublet of bands ($\mu_0=\mu_1=250.356$ meV), the value of $M_x$ driving the topological transition is of the order of $0.05$ meV (thus about one Tesla), with a TSC phase that can be stabilized for an electron density of the order of $10^{14} cm^{-2}$ which is then experimentally accessible both by electric gating and application of an applied magnetic field. Let us note that the realization of TSC requires a fine tuning of the parameters of the Hamiltonian since one has, first, to stabilize the SC phase by fixing the chemical potential close to the peaks of the density of states (DOS) and then vary the magnetic field to achieve a gap closing at $k_x=0$. 


In order to further investigate the consequences of the lateral confinement in setting the topological superconducting phases, we consider nanowires with increasing thickness. 
We find that the number of stability spots for topological superconductivity increases with $N_y$. 
In Fig. \ref{Fig3}, we show the DOS for quasi-one-dimensional nanowires up having $N_y$ equal to $8$ and 14 indicating with red dash lines the electron filling regions where TSC takes place. 


The results for large values of $N_y$ are quite interesting. As discussed in the beginning of this section, in the normal state, starting from $N_y=10$, the bands with $xy$ character are at lower energies than the $xz$ and $yz$ as it occurs in the two-dimensional bulk. Therefore, the sub-bands are not isolated, and, as consequence, fully coupled. We have checked by calculating the invariant $Q$ that all the minima of the sub-bands become spots for topological superconductivity. Therefore, as shown in Fig. \ref{Fig3}(c) for $N_y=14$, there are a lot of spots in the energy window between $0$ and the maximum of the DOS close to $80$ meV. We remark that this maximum is close to the van Hove singularity of the two dimensional bulk\cite{Zhong}. Moreover, as shown in Fig. \ref{Fig3}(d), for this range of energies, the particle density is of the order of $10^{13} cm^{-2}$, which corresponds to the optimal doping value for the bulk superconductivity. 
A further scaling of the nanowire thickness confirms the stability of the topological superconducting phases in the middle energy range, where the sparse distribution of islands (spots) with non-trivial topological ground states is quite stable and reliable up to a length of the order of the superconducting coherence length (in Eq. (\ref{coherence}) $\xi$ has been estimated to be of the order of $80$ nm) where the system retains its quasi-one-dimensional character.

It is interesting to analyze the phase diagrams magnetic field ($M_x$) - filling ($\epsilon_0$) for larger numbers of lateral chains. In Fig. (\ref{Fig5}), we consider the case $N_y=8$ showing the phase diagrams corresponding to two chemical potentials, $\mu_3$ and $\mu_4$, reported in Fig. (\ref{Fig4}). As discussed above, the chemical potential $\mu_3$ marks the minimum of a band with prominent $xy$ character. Therefore, in panel (a) of Fig. (\ref{Fig5}),
the phase diagram bears a strong resemblance with the diagram presented in the panel (c) of Fig. (\ref{Fig3}), where the same $xy$ band is occupied in the case of two chains.  More interesting is the plot shown in the panel (b) of Fig. (\ref{Fig5}). Here, the heavy band $yz$ is occupied, therefore, the stability of the superconducting phases gets enhanced. Indeed, there is a marked increases of the filling energy where the superconductivity is stable. However, the transition to the topological superconducting phase requires a larger value of the magnetic field. Indeed, as shown n the panel (b) of Fig. (\ref{Fig5}), the topological phase becomes accessible only by applying a huge magnetic field. 
Ultimately, we have checked that, in most cases where the band with $yz$ or $zx$ character are the highest to be occupied, one enforces the superconducting phase, but one needs very large magnetic fields to enter the topological regime.  

Finally, we discuss the effect of the magnitude of the pairing energy $U$ on the stability of superconducting phases and the topological transition. Actually, starting from nanowires with a few chains, the topological superconducting phase is achieved when the pairing energy is not large.  Actually, $U$ has not to be small otherwise the superconductivity is not stable. However, $U$ has not to be very large otherwise it overwhelms the magnetic energy hindering the topological regime. As shown in many phase diagrams discussed in this paper, the magnetic energy is typically comparable with the filling energy which, on the other hands, affects the stability of superconducting phases.  There is quite often an intermediate regime of values for $U$ where the topological regime becomes accessible. In particular, we analyze the case $N_y=14$, where, as shown in Fig. (\ref{Fig4}), there are a lot of topological spots in the energy window close to zero. In this window, as shown in the panel (c) of  Fig. (\ref{Fig5}) for a small value of the magnetic field, we have analyzed the stability of superconducting phases for values of $U$ between $120$ meV and $200$ meV. We find that, with increasing the value of the magnetic field, the topological regime is maintained for values of $U$ smaller than $200$ meV, a value where the superconducting phases by different spots start merging. We have found that  this merging turns out to be often detrimental for the stability of topological phases. 



\section{Discussions and Conclusions}

In this paper we have studied the formation and stability of topological superconducting phases in quasi-one-dimensional LAO/STO nano-wires. We have self-consistently solved the Bogoliubov-de Gennes equations for a multi-band electronic model including atomic spin-orbit and orbital Rashba interactions in the presence of a magnetic field lying in the interface plane by evaluating the role of the in-plane confinement due to the finite thickness of the nanowire. 

One of the main results of this paper has been to show that the changeover from sparse-to-dense distribution of topologically non-trivial domains depends  on the multi-band electronic structure and the confinement potential.  
According to the analysis performed in the previous sections, the lateral confinement can completely alter the energy splitting of the Ti $t_{2g}$ states and consequently influences the electron doping distribution of the topological phases. In the regime of strong confinement, the topological states are rare and sparse exhibiting a number of regions which is related to the number of chains forming the nanowire. In any case,  in this confinement regime (up to  ten lateral chains), the number of topological domains is smaller than the number of subbands.  In fact, the topological phase cannot be achieved in the lowest energy orbital band which is insensitive to the lateral confinement. This is due to the fact that the effective spin Rashba coupling is very tiny in those states as it is obtained by virtual inter-band processes which require to overcome a gap of the order of the inter-chain hopping. Once the $xy$ band and the quasi flat $yz$ bands start to get populated, then a more robust topological phase can be settled.  Actually, the increase of the nanowire thickness drives a transition at low energy with an orbital population inversion with the $xy$ band being the lowest in energy occupied orbital configuration. As discussed in this paper, such regime is obtained for LAO-STO nanowires having a thickness of the order of 10 nm. Actually, in the regime of weak confinement (more than ten lateral chains), the number of topological domains scales with the number of subbands. Since LAO-STO nanowires with lateral lengths starting from $5$ nm have already been fabricated \cite{Pai2}, the results found in this paper can be experimentally verified in setups which are accessible with nanofabrication techniques. We stress that all the results  are valid for quasi one-dimensional nanowires, therefore up to lateral widths of the order  the  superconducting coherence length.

At this point it is valuable to compare the microscopic conditions related to the topological superconducting phase in LAO/STO nanowires and that one obtained in InSb or InAs semiconductors. The charge carrier mass in LAO/STO is at least one order of magnitude larger than the effective mass $m^*$ of the above mentioned semiconductors \cite{Lutchyn2}. This is important when multiple sub-bands are formed due to the lateral confinement. Indeed, for smaller effective masses, one would expect a larger energy separation between the sub-bands. Therefore, as found in experiments with semiconductors, the limit of a single sub-band is reachable. Moreover, even if the Rashba coupling constants are different, the Rashba energies are of the same order of magnitude in these systems. Another relevant difference between these systems is in the amplitude of the $g$-factor. The large gyromagnetic factor in semiconductors allows to open a large Zeeman gap using in-plane magnetic fields quite below the critical field of the superconductor.  Even if the $g$-factor in LAO/STO is of the order of two, the self-consistent analysis made in this paper has shown that the in-plane magnetic fields for the topological transition can be of the order of one Tesla, which is still smaller than the critical field of the superconducting phase at optimal doping  \cite{Shalom}.  

Summarizing, some of the characteristics of LAO/STO, like the $g$-factor and the effective mass, are less favorable for obtaining one-dimensional topological superconductivity.
However, beside some disadvantages, according to known electronic and superconducting properties of the LAO/STO system, there are also important advantages over complex hybrid superconducting/semiconducting technology. 
First of all, LAO/STO is a 2D superconductor characterized by large spin-orbit coupling. Thus, in LAO/STO there is no need to interface the material with other superconductors to get proximity effect induced topological superconductivity. This is a clear technological advantage, together with the related  possibility to realize, with a top-down approach, complex device geometries. 
Additionally, LAO/STO is extremely sensitive to gate voltages. The system can locally be tuned from a superconducting to a metallic and even insulating state with small gate voltages. Moreover, as discussed in the paper, due to the 3d-orbital degree of freedom and the fact that the spin-orbit coupling can deviate from the Rashba coupling, it is important to understand which type of topological superconductivity takes place.

Finally, it is worth pointing out that the orientation of the magnetic field in the interface plane plays a crucial role in the stability of the topological phase diagram. We have verified that the application of a magnetic field along the transverse direction of the nanowire is generally detrimental for the occurrence of topological states in the parameters space because it is collinear to the spin polarization induced by the effective Rashba coupling. In this paper, we have always considered the effects of homogeneous magnetic fields on a clean system. Indeed, the study of one-dimensional phenomena due to the coupling of itinerant electrons  with magnetic impurities \cite{Fidkowski2} or the presence of a superconductor/ferromagnet interface \cite{Natcomm} is out of the scope of this paper.  Here, the focus has been on quasi-one-dimensional nanowires at zero temperature in the limit of infinite length analyzing the superconducting properties with increasing the lateral width. Therefore, the effects of fluctuations beyond mean-field should be attenuated in these physical conditions. Even in the strict one-dimensional case, it has been shown that the effect of fluctuations does not destroy the Majorana excitations at the edges \cite{Fidkowski1}. More precisely, a wire with algebraically decaying superconducting fluctuations supports Majorana fermion zero modes and its topological degeneracy can decay as a power law of the length of the superconducting wire \cite{Fidkowski1}. Therefore, due to the bulk-boundary correspondence,  topological phases are expected to be quite robust to fluctuation effects. We believe that the Bogoliubov-De Gennes formalism used in this paper is able to capture at least qualitatively the topological phase transitions and to elucidate the stability of superconducting topological phases.

\section*{Acknowledgements}
The authors acknowledge funding from the project QUANTOX (QUANtum Technologies with 2D-OXides) of QuantERA ERA-NET Cofund in Quantum Technologies
(Grant Agreement N. 731473) implemented within the European Union's Horizon 2020 Programme. The authors  acknowledge stimulating discussions with A. Akhmerov.

\begin{appendix}

\section{One chain nanowire}
\label{AppA}

In this Appendix we present the evolution of the order parameter for a single chain nanowire and the corresponding topological phase diagram.
Although in the single chain limit there are no sub-bands due to the lateral confinement, the solution for the order parameter is representative of the general trend that one obtains when considering the effect of the magnetic field. 

\begin{figure}[t]
\centering
\includegraphics[width=7.0cm,height=9.0cm,angle=-90]{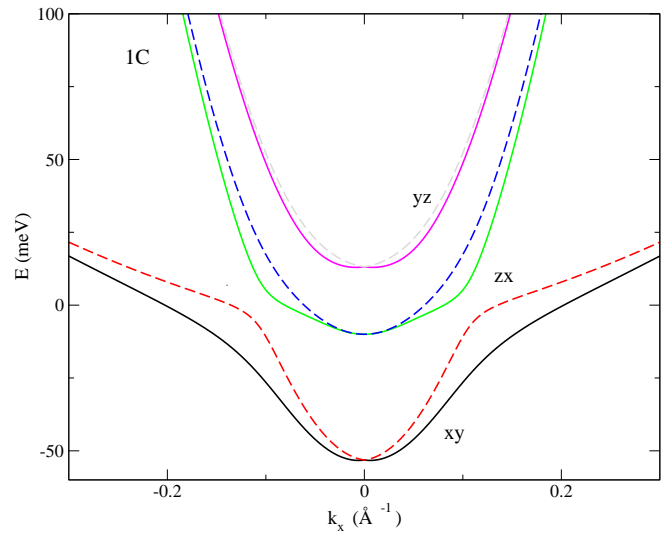}
\caption{Electronic structure of the oxide nanowire for the case of 1-chain (1C).}
\label{Fig100}
\end{figure}

Even in the one-dimensional limit, the spectrum of the normal state is characterized by several bands which cannot be easily compared with those in nano-wires with only Rashba coupling or in nanostructures of topological insulators \cite{Paolino,Marigliano1,Marigliano2,Marigliano3,Perroni1}. Actually, the case with only one chain represents a limiting configuration with vanishing inter-chain charge transfer (see Fig. \ref{Fig100}). In this case, the $xy$ band is the lowest in energy and it exhibits a splitting of the order of $\Delta_t$ with the $zx,yz$ states, that are in turn separated in energy as a consequence of the atomic spin-orbit coupling $\Delta_{SO}$.  For the single chain, the dispersion along the $x$ direction is the same that one would have obtained for the bulk along the high symmetry line from the center of the Brillouin zone ($\Gamma=[0,0]$) to the zone boundary ($X=[\pi,0]$).   This occurs because the electronic structure has only nearest neighbor hopping along x-y directions and inversion asymmetric terms with $sin[k_x a ]$ and $sin[k_y a]$ dependence. Since for the $\Gamma-X$ scan, $k_y=0$, then the corresponding contribution is just a constant shift/offset for the energy and, thus, it does not affect the dispersion. Therefore,  on one hand, the single-chain case is not physically realizable, on the other it is useful for sake of comparison with the more physical quasi-1D case. Finally, we note that lateral confining potential acts to renormalize the energy position of the $yz$ and $xy$ bands. Since this effect can be also mimicked by a suitable renormalization of the lateral hopping, one can directly predict the consequences of the lateral potential on the basis of the obtained results.

\begin{figure}[t]
\centering
\includegraphics[width=10.0cm,height=9.1cm,angle=-90]{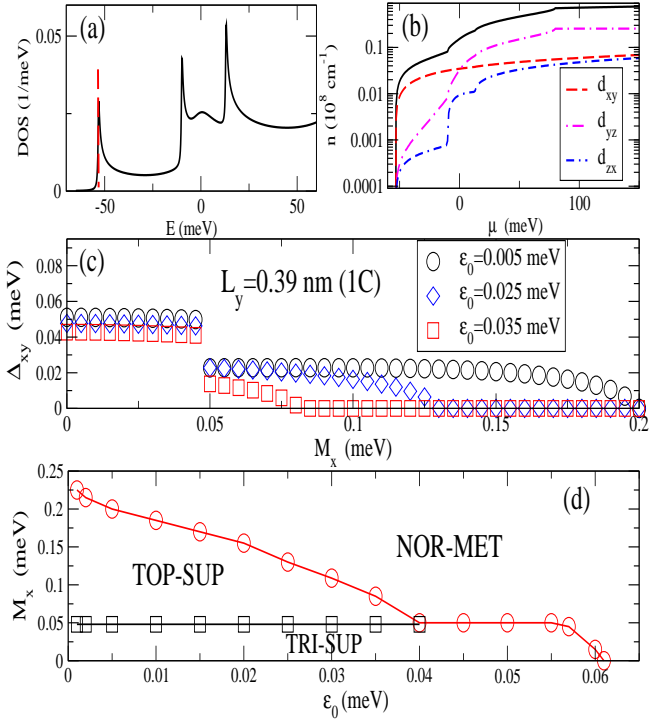}
\caption{(a) Normal state density of states for the $N_y=1$ nanowire. Dotted lines schematically indicate the filling regimes for which a topological superconducting phase can be stabilized. (b) Evolution of the orbital dependent electron density as a function of the energy. The lowest bands have a $xt$ character whereas the $xz$ and $yz$ electronic states start to get populated above a given energy which is set by the crystal field potential $\Delta_t$.  (c) Evolution of the superconducting order parameter as a function of the applied magnetic field along the nanowire long axis direction for different value of the offset energy $\epsilon_0$ for the intra-band filling control. (d) Phase diagram at a given value of the band minimum chemical potential in terms of the electron filling offset $\epsilon_0$ and the applied magnetic field $M_x$ along the $x$ direction of the nanowire. TOP-SUP, TRI-SUP and NOR-MET stand for topological superconducting, trivial superconducting, and normal metallic phase, respectively.}
\label{FigAppA}
\end{figure}

The density of states in the absence of superconductivity is reported in Fig. \ref{FigAppA}(a).  Dotted lines in Fig. \ref{FigAppA}(a) indicate that, for a single-chain wire, a topological superconducting phase can be stabilized only at low energies where the electronic states are described by $xy$-like bands.

In Fig. \ref{FigAppA}(b), we report the evolution of the order parameter obtained by self-consistent iterative procedure.
The self-consistent calculation of $\Delta_{xy}$ points out that the order parameter  is nearly flat before and after the transition, it exhibits a jump of $0.025$ meV at the transition ($M_x=0.0456$ meV in Fig. \ref{Fig3}), and, finally, it decreases towards zero at larger magnetic fields. Therefore, the fingerprint of the transition is already found in the behavior of the order parameter confirming that Rashba and magnetic energies are of the same order of magnitude\cite{Tewari}.

In order to induce the transition into the topological state, not only the gap but also the energy $\epsilon_0$ has to be small (it is set equal to $0.005$ meV in Fig. \ref{FigAppA}). Actually, if one defines the critical energy $M_{c3}$ as
\begin{equation}
M_{c3}=\sqrt{\Delta_{xy}^2+\epsilon_0^2}, 
\label{emmec3}
\end{equation}
the topological transition takes place when the magnetic energy $M_x$ overcomes $M_{c3}$ \cite{Aguado1}

In order to pursue the analysis in the regime of low density, we study the range of energies $\epsilon_0$ where topological superconductivity is observed. Since 
$\epsilon_0$ appears in the critical energy $M_{c3}$ defined in Eq. (\ref{emmec3}), one expects that these values of $\epsilon_0$ cannot be large. Actually, they have to be smaller than the gap  for experimentally accessible values of the magnetic energies $M_x$.  As shown Fig. \ref{FigAppA}, we have studied the order parameter $\Delta_{xy}$ as a function of the magnetic energy $M_x$ for different values of $\epsilon_0$. We point out that, as expected for one dimensional systems where the van Hove singularity in the density of states (DOS) is at the bottom of the band, an increase of the density induces a weakening of the superconducting phase. Therefore, the magnitude of the order parameter decreases with increasing the filling energy $\epsilon_0$ for all the values of the magnetic energy $M_x$. However, we point out the rapid decrease of the order parameter at the transition for larger values of the filling energy $\epsilon_0$. Moreover, the energy position of the topological transition is practically constant with varying $\epsilon_0$ ($M_x$ close to $0.05$ meV in Fig. \ref{FigAppA}). Therefore, as shown in Fig. \ref{FigAppA}, the transition line between trivial and topological superconductivity is straight. Furthermore, in the phase diagram, we notice a range of higher densities where the transition from the  superconductor to the normal metal occurs directly from the trivial superconducting state.  The general trend is a reduction of the superconductivity with increasing the density in the regime where the pairing energies are not large.

\section{Determination of the topological invariant}   
\label{AppB}

In this Appendix, we provide some details to calculate the Pfaffian invariant $Q$ defined in Eq. (\ref{invariant}).  
For simplicity, we consider the case of one-dimensional wires. As discussed in previous sections, the same calculation procedure for $Q$ will be also used in the case of quasi-one-dimensional nanowires.

First, one needs to introduce the  couple $\gamma_{k_x,r,\sigma,a}$ and 
$\gamma_{k_x,r,\sigma,b}$  of Majorana fermions at fixed wave-vector $k_x$, band index $r$ and spin $\sigma$:
\begin{eqnarray}
c_{k_x,r,\sigma} & =& \frac{1}{2} (\gamma_{k_x,r,\sigma,a}-i \gamma_{k_x,r,\sigma,b})   \nonumber \\
c_{-k_x,r,\sigma}^{\dagger} & =& \frac{1}{2} (\gamma_{k_x,r,\sigma,a}+i \gamma_{k_x,r,\sigma,b}),
\label{hamilbogo}
\end{eqnarray}
where $c_{k_x,r,\sigma}$ ($c_{-k_x,r,\sigma}^{\dagger}$) indicates the annihilation (creation) operator of electrons with spin $\sigma$ relative to the orbital $r$ at the wave-vector $k_x$ ($-k_x$).
The spinor $\underline{\psi}_{k_x} $ with $12$ electronic operators can be transformed in term of the spinor  $\underline{\gamma}_{k_x} $ defined in terms of $12$ Majorana fermions
such that 
\begin{equation}
\underline{\psi}_{k_x} = C \cdot \underline{\gamma}_{k_x} 
\end{equation}
with $C$ the following $12 \times 12$ matrix 
\[
C= \\
\left(
{ \begin{array}{ccc}
  C_{1,1} & C_{1,2} & C_{1,3}  \\
  C_{2,1} & C_{2,2} & C_{2,3}  \\
  C_{3,1} & C_{3,2} & C_{3,3} 
 \end{array} } \right),
\]  
where the  $4\times4$ sub-matrices $ C_{i,j}$, with $i,j=1,2,3$, are the following:
\[
 C_{i,i}= \frac{1}{2}\\
\left(
{ \begin{array}{cccc}
  1 & -i & 0 & 0\\
  0 & 0 & 1 & -i  \\
  1& i & 0 & 0   \\
  0 & 0 & 1& i 
\end{array} } \right),
\]  
with $0=C_{1,2}=C_{2,1}=C_{1,3}=C_{3,1}=C_{2,3}=C_{3,2}$. Finally, one can define the real $12 \times 12$ matrix  $B(k_x)$ such that
\begin{equation}
i B(k_x) =C^{\dagger} \cdot K_{BdG}(k_x) \cdot C,
\label{invariant1}
\end{equation}
with $K_{BdG}(k_x)$ Bogoliubov-de Gennes matrix  used for the definition of the hamiltonian operator at mean-field level. For the calculation of  the Pfaffian invariant defined in Eq. (\ref{invariant}), one has to calculate the Pfaffian of the skew-symmetric matrix $B(k_x)$ at $k_x=0$ and $k_x=\pi/a$. In the one-dimensional case, the calculation of the Pfaffian of a $12 \times 12$ matrix can be made directly from its definition. However, in order to calculate the Pfaffian also in the case of quasi-one-dimensional  wires, we implement its numerical calculation by using the Hessenberg decomposition  of the matrix $B(k_x)$:
\begin{equation}
B(k_x) =U(k_x) F(k_x)  U(k_x)^{T},
\label{invariant2}
\end{equation}
where $U(k_x)$ is orthogonal and $F(k_x)$ is an upper Hessenberg matrix, meaning that it has zeros below the first sub-diagonal \cite{Wimmer2}. In the case of $k_x=0$ and $k_x=\pi/a$,  the upper Hessenberg matrix F  not only is skew-symmetric, but it  is also tridiagonal. Therefore, one can use the following property of the Pfaffian under orthogonal transformations:
\begin{equation}
P \left[ B\left(k_x=0/\frac{\pi}{a}\right) \right]=det\left[ U\left( k_x \right) \right] P\left[ F\left(k_x=0/\frac{\pi}{a}\right) \right],
\label{invariant3}
\end{equation}
where the determinant of $U(k_x)$ is $1$ or $-1$. The calculation of the Pfaffian of $F(k_x=0/\pi/a)$ is simple, since it is given by the product of the odd elements of the first upper sub-diagonal.

\end{appendix}

\end{document}